\documentclass[aps,prl,twocolumn,superscriptaddress,amssymb]{revtex4}
\usepackage{graphicx}
\usepackage{amsmath}
\usepackage{amsfonts}
\usepackage{epsfig}
\usepackage{bm}
\usepackage{color}

\begin{document}

%\title{Superconductivity in the doped Hubbard model on four-leg cylinders}
\title{Superconductivity in the doped Hubbard model and its interplay with charge stripes and next-nearest hopping $t^{\prime}$}
\author{Hong-Chen Jiang}
\email{hcjiang@stanford.edu}
\affiliation{Stanford Institute for Materials and Energy Sciences, SLAC and Stanford University, Menlo Park, California 94025, USA}
\author{Thomas P. Devereaux}
\affiliation{Stanford Institute for Materials and Energy Sciences, SLAC and Stanford University, Menlo Park, California 94025, USA}
\affiliation{Geballe Lab for Advanced Materials, Stanford University, Stanford, California 94305, USA}
\date{\today}

%%%%%%%%%%%%%%%%%%%%%%%%%%%%%%%%%%%%%%%%
\begin{abstract}
We report a large-scale density-matrix renormalization group study of the lightly doped Hubbard model on 4-leg cylinders at hole doping concentration $\delta=12.5\%$. By keeping a large number of states for long system sizes, we are able to reveal a delicate interplay between superconductivity and charge and spin density wave orders tunable via next-nearest neighbor hopping $t^{\prime}$. For finite $t^\prime$, the ground state is consistent with that of a Luther-Emery liquid, having ``half-filled'' charge stripes with power-law superconducting and charge-density-wave correlations of wave-length $\lambda=1/2\delta$, but short-range spin correlations. This is in direct contrast to the case with $t^{\prime}=0$, where superconducting correlations fall off exponentially while charge- and spin-density modulations are dominant. Our results indicate that a route to robust long-range superconductivity involves destabilizing insulating charge stripes in the doped Hubbard model.
\end{abstract}
\maketitle

%%===============Introduction======================
%\textbf{Introduction:} %
Despite intense numerical studies of the two-dimensional (2D) Hubbard model, the critical question of whether the model supports the presence of robust superconducting order remains unclear, owing in part to the close competition between a number of near-degenerate ground states composed of various electronic orders \cite{Corboz}. While finite temperature studies using cluster dynamical mean field theory indicate a transition into a uniform d-wave superconducting state at temperatures $\sim 0.02 t$ \cite{Staar}, extensive studies using density matrix renormalization group (DMRG), particularly around $\delta=12.5\%$ doping, indicate the charge and spin density wave order in the form of "stripes" provide dominant correlations, with superconducting correlations being subleading and decaying exponentially with lattice size \cite{Zheng_Science_1155,Huang_Science_1161,Ehlers_PRB_2017}. In addition it was shown that the filling of the stripes and the wavelength of them depends strongly on next-nearest neighbor hopping $t^{\prime}$, with a much smaller dependence on Hubbard $U$, reflecting a flat energy landscape for the way in which stripes can appear in Hubbard ladders \cite{Zheng_Science_1155,Ehlers_PRB_2017,Huang2}. This raises the intriguing possibility, due to the delicate interplay between stripe and superconducting order, that the underlying superconducting state of the Hubbard model might be quite sensitive to $t^{\prime}$, as discussed empirically and in the context of the role of "axial orbitals" \cite{Pavarini}.

In this paper we report extensive DMRG studies of the $t-t^{\prime}-U$ Hubbard model at hole doping concentration $\delta=12.5\%$ on 4-leg cylinders with periodic and open boundary condition in short and long directions, respectively. By explicitly keeping a large number of states, we demonstrate that the equal-time superconducting (SC) and charge density wave (CDW) correlations decay with power laws. Consistent with Luther-Emery (LE) liquid\cite{lutheremery}, this demonstrates a close interplay between charge and superconducting correlations. Moreover, we show that the results depend strongly on $t^{\prime}$, which tips the balance between charge density and superconducting correlations. Specifically, we find ``filled'' insulating charge stripes of wavelength $\lambda=1/\delta$ and a lack of long-range superconductivity for $t^\prime=0$, which is consistent with prior results.\cite{Zheng_Science_1155,Ehlers_PRB_2017} For finite $t^\prime<0$, the insulating charge stripes are replaced with weaker "half-filled" stripes with a shorter period $\lambda=1/2\delta$, and concomitantly, the superconducting correlations become long-ranged. As far as we know, our results are the first demonstration of LE liquid in Hubbard systems with long-range superconducting correlations on cylinders or ladders wider than 2, with a delicate interplay with doping of charge stripes and superconductivity modified solely by $t^{\prime}$.

%=============Fig1: Charge density profile===============
\begin{figure}
  \includegraphics[width=\linewidth]{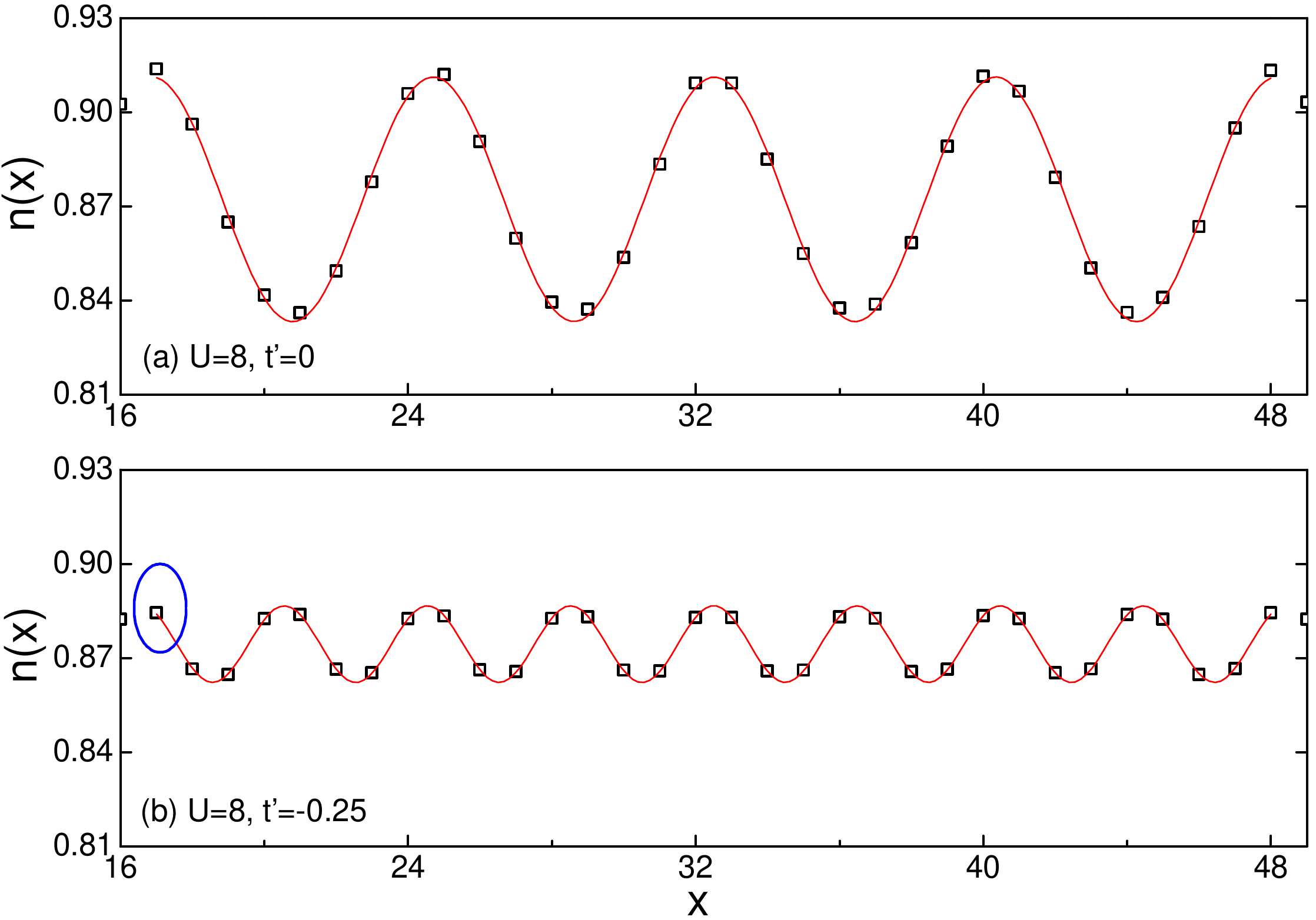}
  \caption{(Color online) Charge density profile $n(x)$ for the Hubbard model at doping level $\delta=12.5\%$ on $L_x=64$ cylinders at $U=8$ with $m=20000$ for (a) $t^\prime=0$ and (b) $t^\prime=-0.25$. The black squares denote numerical data, while the red lines are fitting curves using function $n(x)=A_{cdw} \cos (Q x + \theta) +n_0$, where $A_{cdw}$ and $Q$ are the CDW amplitude and ordering wavevector, respectively. Note that only the central-half region with rung indices $x=\frac{L_x}{4}+1\sim \frac{3L_x}{4}$ are shown and used in the fitting while the remaining $\frac{L_x}{4}$ data points from each end are removed to minimize boundary effects. The blue oval labels the ``reference site'' (see text).}\label{Fig:DensityProfile}
\end{figure}

%%==========Model and Method=============
\textbf{Model and Method:} We employ DMRG \cite{White_PRL_1992} to investigate the ground state properties of the hole-doped Hubbard model on the square lattice defined by the Hamiltonian%
\begin{eqnarray}\label{Eq:ModelHamiltonian}
H =- \sum_{ij\sigma} t_{ij} \left(\hat{c}^+_{i\sigma} \hat{c}_{j\sigma} + h.c.\right)+ U\sum_i \hat{n}_{i\uparrow}\hat{n}_{i\downarrow}, 
\end{eqnarray}
where $\hat{c}^+_{i\sigma}$ ($\hat{c}_{i\sigma}$) is the electron creation (annihilation) operator on site $i=(x_i,y_i)$ with spin $\sigma$, $\hat{n}_i=\sum_{\sigma}\hat{c}^+_{i\sigma}\hat{c}_{i\sigma}$ is the electron number operator. The electron hopping amplitude $t_{ij}$ is equal to $t$ if $i$ and $j$ are nearest-neighbors (NN) and equal to $t^\prime$ for next-nearest-neighbors (NNN). $U$ is the on-site repulsive Coulomb interaction. We take the lattice geometry to be cylindrical and a lattice spacing of unity. The boundary condition of the cylinders is periodic in the $\hat{y}=(0,1)$ direction while open in the $\hat{x}=(1,0)$ direction. Here, we focus on cylinders with width $L_y$ and length $L_x$, where $L_y$ and $L_x$ are number of sites along the $\hat{y}$ and $\hat{x}$ directions, respectively. There are $N=L_x\times L_y$ lattice sites and the number of electrons is $N_e=N$ at half-filling, i.e., $\hat{n}_i=1$. The concentration of doped holes is defined as $\delta=\frac{N_h}{N}$ with $N_h=N-N_e$ the number of holes which is $N_h=0$ at half-filling.

For the present study, we focus on the lightly doped case at hole concentration $\delta=12.5\%$ on cylinders with width $L_y=4$ and length up to $L_x=64$. We set $t=1$ as an energy unit and report results for $t^\prime=-0.25$ with interactions $U=8$ and $U=12$. For comparison, the case with $t^\prime=0$ is also considered. In our calculations, the total magnetization is fixed at zero and we perform around 60 sweeps and keep up to $m=20000$ number of states in each DMRG block with a typical truncation error $\epsilon\sim 1\times 10^{-6}$ for $t^\prime=-0.25$ and $\epsilon\sim 3\times 10^{-6}$ for $t^\prime=0$. This leads to excellent convergence for our results when extrapolated to $m=\infty$ limit. Further details of the numerical simulation are provided in the Supplemental Material.

%%===============Fig2: CDW order parameter===============
\begin{figure}
  \includegraphics[width=\linewidth]{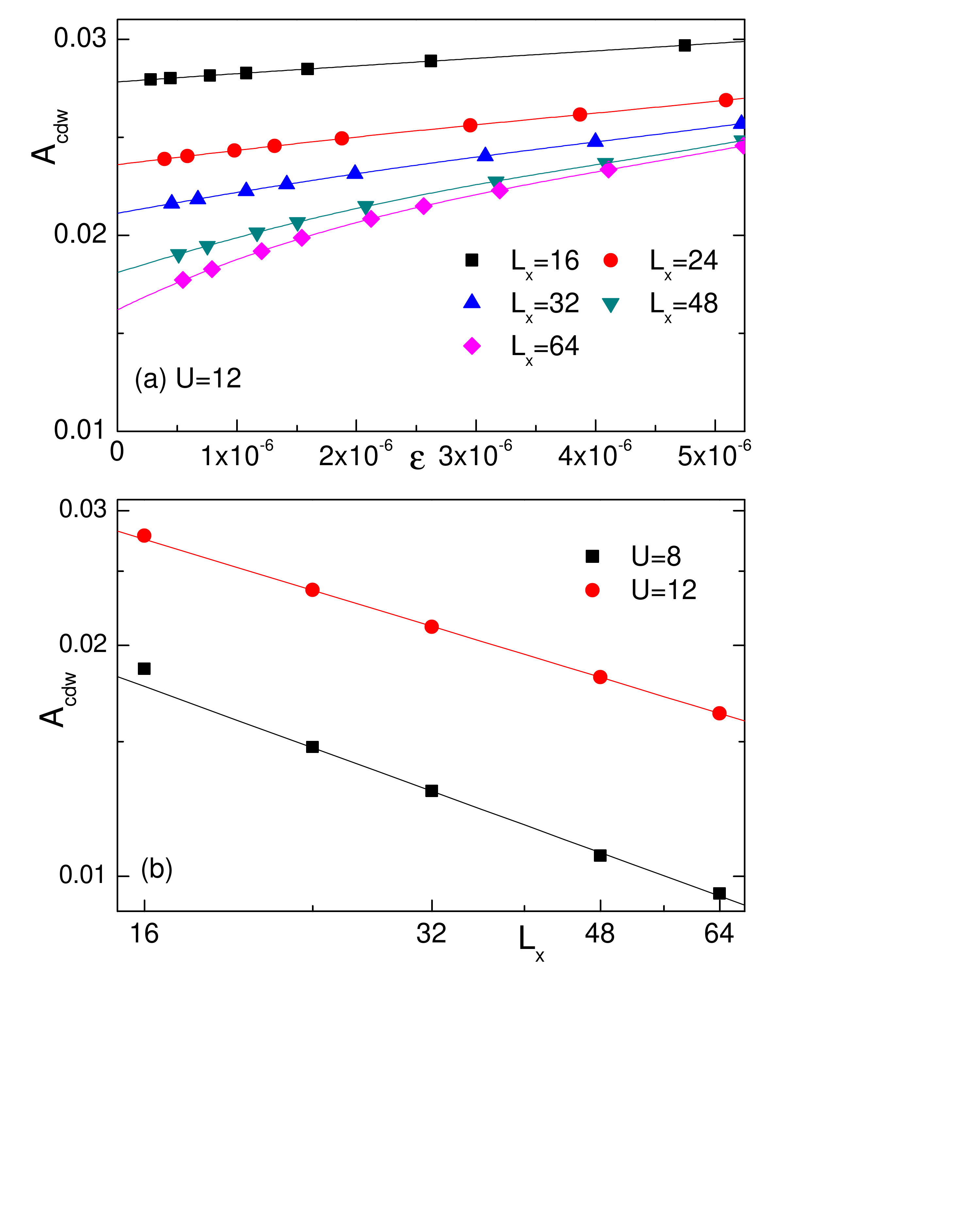}
  \caption{(Color online) (a) Convergence and length dependence of the CDW amplitude $A_{cdw}(L_x)$ for $\delta=12.5\%$ at $t^\prime=-0.25$ and $U=12$ with $L_x=16\sim 64$. The solid lines denote fittings using quartic polynomials. (b) Finite-size scaling: $A_{cdw}(L_x)$ as a function of $L_x$ in a double-logarithmic plot for both $U=8$ and $U=12$.}\label{Fig:CDW}
\end{figure}

%%============Principle results===============
\textbf{Principal results:} %
We have investigated the ground state properties of the Hubbard model on a $L_y=4$ cylinder at doping level $\delta=12.5\%$ with interaction $U=8$ and $U=12$. For $t^\prime=0$, we find that the system forms charge stripes of wavelength $\lambda=1/\delta$, i.e., $\lambda=8$, and antiferromagnetic ordering with a modulation of wavelength $\lambda=2/\delta$, i.e., $\lambda=16$. Consistent with Hartree-Fock calculations\cite{zaanen,machida,schulz} and previous numerical studies\cite{Ehlers_PRB_2017,Zheng_Science_1155}, these charge stripes carry a wavevector $Q=2\pi\delta$ and so there is one doped hole per unit cell which is referred to as ``filled'' stripes. However, we find that these ``filled'' stripes are not stable with respect to small finite $t^\prime$, where $t^\prime=-0.25$ is enough to drive the system into a new type of charge stripe. Different with the ``filled'' stripes, the new stripes carry an ordering vector $Q=4\pi\delta$ of wavelength $\lambda=1/2\delta$, i.e., $\lambda=4$, with only half a doped hole per unit cell - what is referred to as ``half-filled'' stripes. We have also obtained similar results for other values of $U$ and $t^\prime$.

Thought of as a one-dimensional (1D) system as $L_x\gg L_y$, we find that the ground state is always in a LE phase \cite{lutheremery}, which is characterized by one gapless charge mode but with a gap in the spin sector. The spatial decay of the charge density correlation $A_{cdw}(L_x)$ and the superconducting pair-field correlation $\Phi(r)$ defined in Eq.(\ref{Eq:SCOR}) at long distance are dominated by a power-law with appropriate exponents $K_c$ and $K_{sc}$ defined by%
\begin{eqnarray}
A_{cdw}(L_x)\propto L_x^{-K_c/2} \ {\rm and} \ %
\Phi(x)\propto |x|^{-K_{sc}}, \label{Eq:Kc}
\end{eqnarray}
where $x$ is the displacement along the cylinder $1\ll |x| \ll L_x$. As expected theoretically from the LE liquid, we find the relation $K_c K_{sc}=1$ holds within the numerical uncertainty.  This is in sharp contrast to previous studies \cite{Ehlers_PRB_2017,Zheng_Science_1155} without NNN electron hopping term, i.e., $t^\prime=0$, where the ``filled'' stripes persist in the limit $L_x=\infty$ while a quasi-long-range superconducting correlation is absent. It is however consistent with recent DMRG results from the lightly doped $t$-$J$ model on 4-leg cylinders with ``half-filled'' charge stripes \cite{Jiang_tJ_4leg}.

As the interaction is decreased from $U=12$ to $U=8$, the CDW correlations become weaker while the superconducting correlations become stronger. These numerical observations indicate that CDW and SC may be mutually competing orders, which is consistent with recent experiments on cuprates. Moreover, calculations of spin-spin correlations show that although it is the dominant correlation at short distance, it decays exponentially with distance, allowing for SC correlations to be dominant at long distances. This short-range antiferromagnetic order with gapped spin excitations for $t^{\prime}<0$ is contrary to the case of $t^\prime=0$ where the spin-spin correlations may be long-ranged \cite{Dodaro_PRB_2017,Zheng_Science_1155}, preventing the growth of SC correlations. Therefore, $t^{\prime}$ clearly is a control parameter that tips a delicate balance between CDW, AF, and SC correlations.

%%============Table: Luttinger exponent===============
\begin{table}[t]
\centering \vspace{2mm}
\begin{tabular}[b]{|c|c|c|c|c|}
\hline \hspace{0mm} $U$ \hspace{4mm}&\hspace{4mm} $K_c$\hspace{4mm} & \hspace{4mm} $K_{sc}$\hspace{4mm} & \hspace{2mm} $K_c K_{sc}$\hspace{4mm} & $\xi_s$\hspace{4mm}\\
\hline$8$&$0.90(6)$ & 1.43(8) & 1.3 (2)& 9.8(6) \\ %\hline
\hline$12$&$0.75(6)$ & 1.60(7) & 1.2(2) & 8.3(4) \\ \hline
\end{tabular}
\caption{List of exponents $K_c$ and $K_{sc}$, and spin-spin correlation length $\xi_s$ of the Hubbard model at doping level $\delta=12.5\%$ and $t^\prime=-0.25$. Here $t=1$.}
\label{Table:KK}
\end{table}

%%=============Charge density wave order==============
\textbf{Charge density wave order:} To describe the charge density properties of the ground state, we define the local rung density operator as $\hat{n}(x)=\frac{1}{L_y}\sum_{y=1}^{L_y}\hat{n}(x,y)$ and its expectation value as $n(x)=\langle \hat{n}(x)\rangle$. Fig. \ref{Fig:DensityProfile}(a) shows the charge density distribution $n(x)$ in a central portion of a cylinder with $L_x=64$ at $U=8$ and $t^\prime=0$, in which the ``filled'' charge stripe of wavelength $\lambda=8$ is found, consistent with previous studies \cite{zaanen,Ehlers_PRB_2017,Zheng_Science_1155,Huang_Science_1161}. The case of $U=12$ gives the similar behavior (not shown). The spin-spin correlations are antiferromagnetic with a $\pi$-phase shift every eight sites as expected. A (quasi)-long-ranged superconducting correlation in this case is unlikely since the charge stripes are completely filled with holes and therefore insulating \cite{Ehlers_PRB_2017} \footnote{We have checked the $t$-$t^\prime$-$J$ model at 1/8 doping and $t/J=3$ and $t^\prime/J=1/4$ with ``filled'' charge striped ground state\cite{Dodaro_PRB_2017}. We have reached to the same conclusion as of the Hubbard model at $t^\prime=0$ where a true long-range CDW ordering is expected.}.
Importantly, we find that a finite $t^\prime=-0.25t$ is sufficient to destroy the insulating charge stripes and the ``half-filled'' stripes of wavelength $\lambda=4$ appear accordingly. An example can be found in Fig. \ref{Fig:DensityProfile}(b), where the charge density modulation $n(x)$ in a central portion of a cylinder with $L_x=64$ at $U=8$ is given. A key feature is that this ``hall-filled'' charge stripe is much weaker than the ``filled'' charge stripe whose modulation amplitude $A_{cdw}(L_x)$ is significantly weaker.

For a given $L_x$ cylinder, the CDW amplitude $A_{cdw}(L_x)$ shown in Fig. \ref{Fig:DensityProfile} can be obtained by extrapolating to the limit $m=\infty$ or $\epsilon=0$. Fig. \ref{Fig:CDW}(a) plots the CDW amplitude $A_{cdw}$ for cylinders of  length $L_x=16\sim 64$ at $U=12$ by keeping $m=4096$ - $20000$ states. It is worth to mention that in the DMRG simulation, accurately describing the behavior of physical observables such as $A_{cdw}(L_x)$ and correlation functions such as $\Phi(x)$ at longer distances requires an increasing number of states and higher order terms in the extrapolation become more important. Therefore, aside from keeping an exceedingly large number of states, we also perform a quartic polynomial fitting to capture the effect of possible higher order terms in the extrapolation. For all cases, we find that this procedure works very well with the linear regression $R^2$ always larger than 99.99\%. Further details concerning the reliability of this extrapolation is presented in the Supplemental Materials.

%The finite-size scaling of $A_{cdw}(L_x)$ can be performed after obtaining its accurate value for given cylinders. 
The results from finite-size scaling of the obtained $A_{cdw}(L_x)$ as a function of $L_x$ are given in Fig.\ref{Fig:CDW}(b) for $t^\prime=-0.25$ at both $U=8$ and $U=12$. In the double-logarithmic plot, our results for both $U=8$ and $U=12$ are approximately linear, indicating that $A_{cdw}(L_x)$ decays with a power-law and vanishes in the limit $L_x=\infty$. The exponent $K_c$, which is shown in Table \ref{Table:KK}, was obtained by fitting the data points using  Eq. (2). $K_c$ can also be obtained directly from the Friedel oscillations of the charge density modulation near the end of the cylinders, giving similar results. Further details can be found in the Supplemental Materials.

%%===============Fig3: SC order parameter===============
\begin{figure}
  \includegraphics[width=\linewidth]{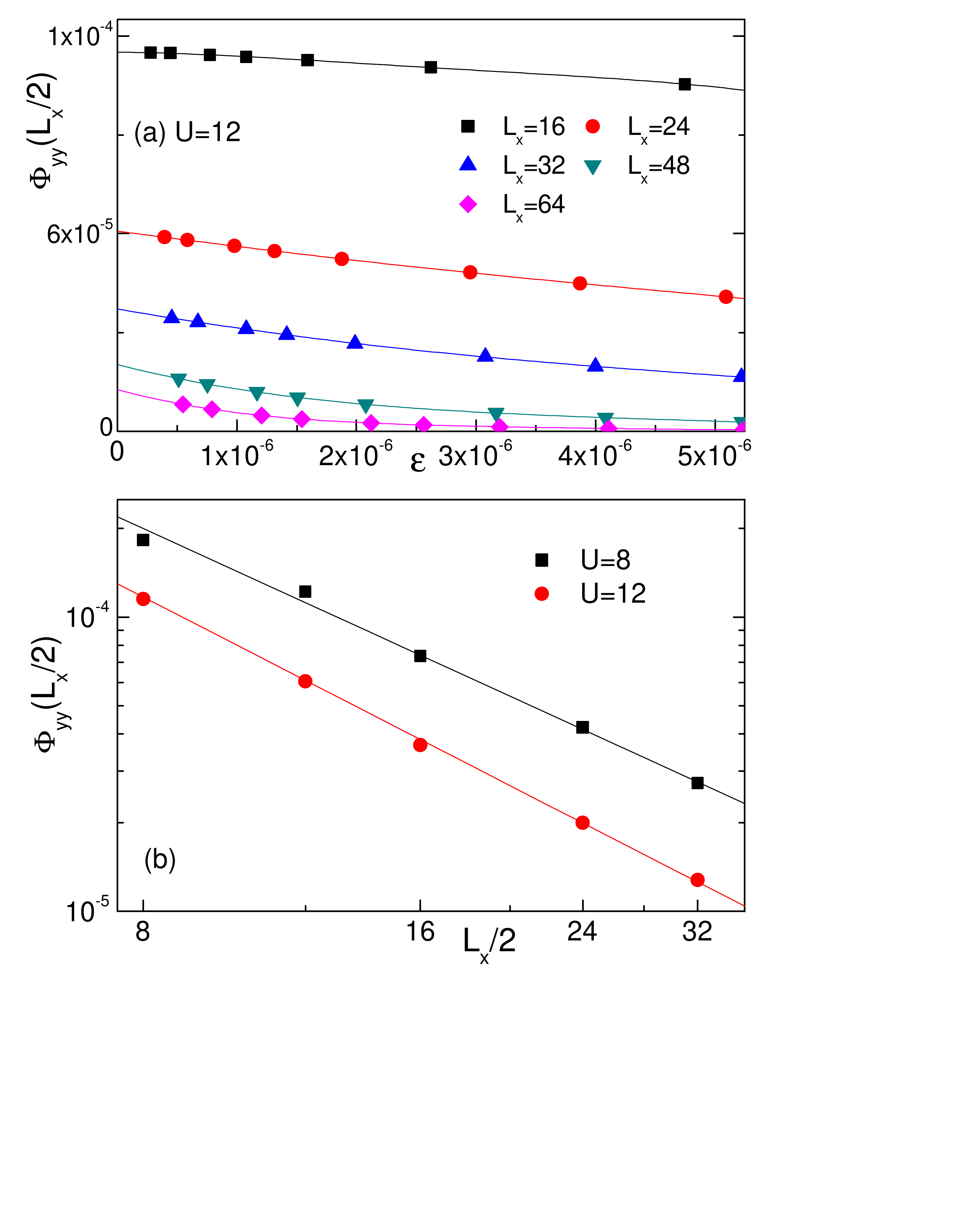}
  \caption{(Color online) (a) SC correlation function $\Phi_{yy}(L_x/2)$ for $\delta=12.5\%$ at $t^\prime=-0.25$ and $U=12$ with $L_x=16\sim 64$. The solid lines denote quartic polynomial fits. (b) Finite-size scaling of $\Phi_{yy}(L_x/2)$ as a function of $L_x$ in a double-logarithmic plot for both $U=8$ and $U=12$.} \label{Fig:SC}
\end{figure}

%%============Superconducting correlation=============
\textbf{Superconducting correlation:} %
In order to test the possibility of superconductivity, we have calculated the equal-time pair-field correlation functions. Since the ground state of the system with even number of doped holes is always found to have spin 0, we focus on spin-singlet pairing. A diagnostic of the SC order is the pair-field correlator, defined as%
\begin{eqnarray}
\Phi_{\alpha\beta}(x)=\frac{1}{L_y}\sum_{y=1}^{L_y}\ \langle \Delta_\alpha^\dagger (x_0,y)\Delta_\beta(x_0+x,y)\rangle.\label{Eq:SCOR}
\end{eqnarray}
Here $\Delta_\alpha^\dagger(x,y)$ is the spin-singlet pair-field creation operator given by $\Delta_\alpha^\dagger(x,y)=\frac{1}{\sqrt{2}}[c_{(x,y),\uparrow}^\dagger c_{(x,y)+\alpha,\downarrow}^\dagger - c_{(x,y),\downarrow}^\dagger c_{(x,y)+\alpha,\uparrow}^\dagger]$, where the bond orientations are designated $\alpha=\hat{x}$, $\hat{y}$, $(x_0,y)$
is the reference bond indicated by the blue oval as shown in Fig. \ref{Fig:DensityProfile}, and $x$ is the distance between two bonds in $(1,0)$ direction.

Due to the presence of CDW modulations (Fig.\ref{Fig:DensityProfile}), SC correlations $\Phi_{\alpha\beta}(x)$ exhibit similar spatial oscillations with $n(x)$. This modulation, together with a significant boundary effect due to open ends of the cylinder, makes it very difficult to accurately determine the decay of SC correlations. This could be one of the main reasons that previous studies have had difficulty in providing direct evidence for (quasi)-long-ranged superconductivity \cite{Dodaro_PRB_2017, Zheng_Science_1155}.

We determine the decay of SC correlations by minimizing the effects induced by both CDW modulations and open boundary conditions simultaneously. Instead of directly fitting $\Phi_{\alpha\beta}(x)$, we calculate the SC correlation $\Phi_{\alpha\beta}(L_x/2)$ for a given cylinder of  length $L_x$, with the reference bond located at the peak position around $x_0\sim L_x/4$ of the charge density distribution. Examples of $\Phi_{yy}(L_x/2)$ are shown in Fig. \ref{Fig:SC}(a) for cylinders of length $L_x=16\sim 64$. Interestingly, we find that the superconducting correlations are much stronger for bonds along the width of cylinder $\Phi_{yy}$ than along the length $\Phi_{xx}$, indicating a possible equal superposition of $d-$wave and extended $s-$wave pairing due the explicit breaking of $C_4$ symmetry on the cylinder.

For each cylinder of length $L_x$, we extrapolate $\Phi_{yy}(L_x/2)$ to the limit $\epsilon=0$ using a quartic polynomial fit with a linear regression $R^2$ larger than $99.97\%$. This gives accurate values of $\Phi_{yy}(L_x/2)$ for reliable finite scaling. More details of the extrapolation are presented in the Supplemental Materials. Fig. \ref{Fig:SC} shows examples of the finite-size scaling of $\Phi_{yy}$ for both $U=8$ and $U=12$. Similar with $A_{cdw}(L_x)$, it also decays with a power-law, whose exponent $K_{sc}$, given in Table.\ref{Table:KK}, was obtained by fitting the results using Eq. (\ref{Eq:Kc}). Therefore, we can conclude that the ground state of the lightly doped Hubbard model at doping level $\delta=12.5\%$ on width $L_y=4$ cylinders with $t^{\prime}=-0.25$ has quasi-long-range SC correlations. This is in stark contrast to the case for $t^{\prime}=0$, where SC correlations decay exponentially and the stripes are filled  \footnote{We have checked the $t$-$t^\prime$-$J$ model at 1/8 doping and $t/J=3$ and $t^\prime/J=1/4$ with ``filled'' charge striped ground state \cite{Dodaro_PRB_2017} and reached to the same conclusion as of the Hubbard model at $t^\prime=0$, where the superconducting correlations decay exponentially.}. 

%===============Fig4:SC and SS===============
\begin{figure}
  \includegraphics[width=\linewidth]{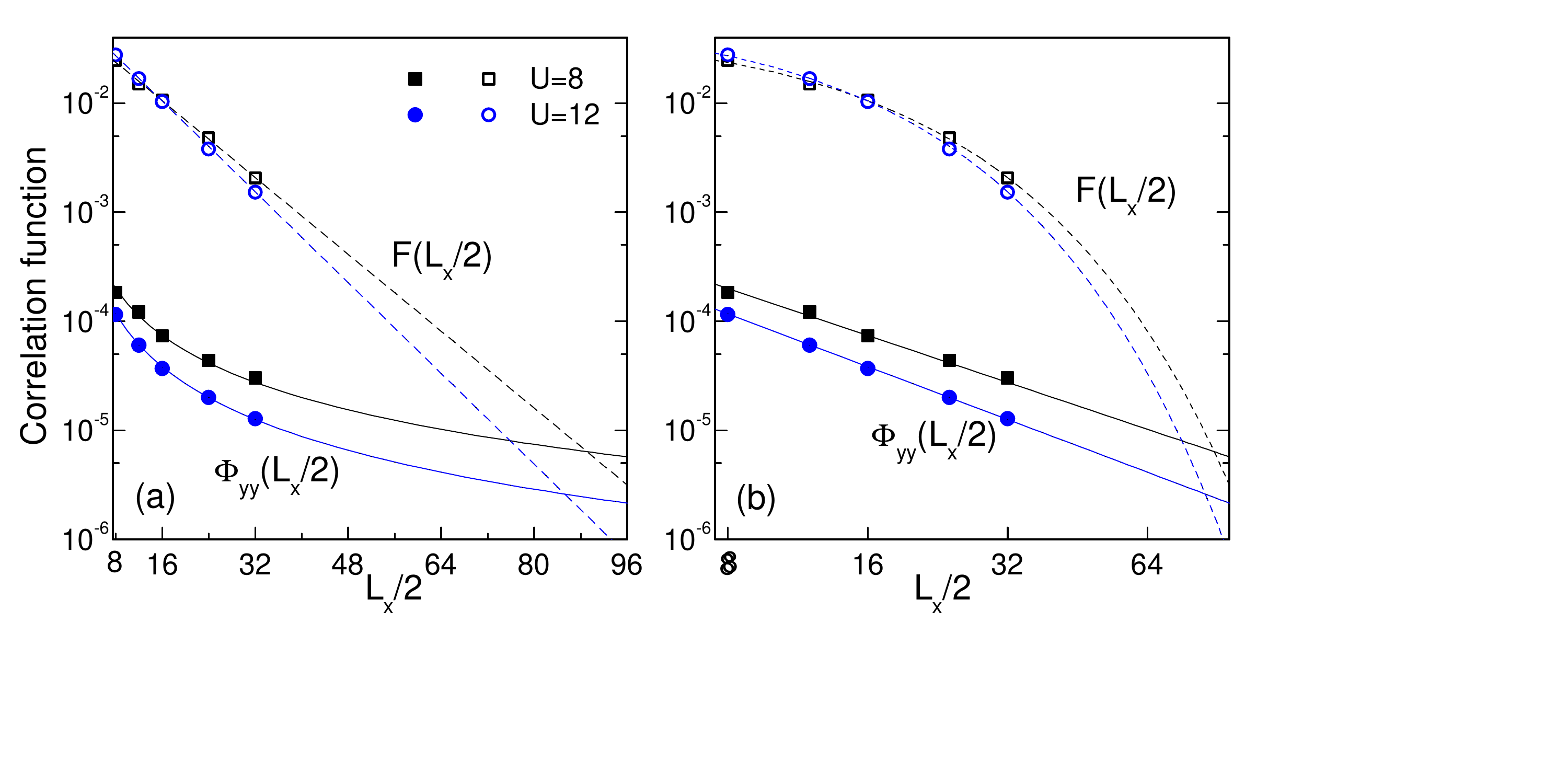}
  \caption{(Color online) SC $\Phi_{yy}(L_x/2)$ and spin-spin $F(L_x/2)$ correlations at doping level $\delta=12.5\%$ at $t^\prime=-0.25$ for $U=8$ and $U=12$ in semi-logarithmic (a) and double-logarithmic (b) scales, as a function of $L_x$. The solid lines denote the power-law fitting $\Phi_{yy}(L_x/2)\propto (L_x/2)^{-K_{sc}}$, while the dashed lines denote the exponential fitting $F(r)\propto e^{-L_x/2\xi_s}$ where $\xi_s$ is the spin-spin correlation length given in Table \ref{Table:KK}.}\label{Fig:SC_SS}
\end{figure}

%%===========Spin-spin correlation==============
\textbf{Spin-spin correlation:} %
To describe the magnetic properties of the ground state, we have also calculated the spin-spin correlation functions defined as%
\begin{eqnarray}\label{Eq:SpinCor}
F(x)=\frac{1}{L_y}\sum_{y=1}^{L_y}|\langle \vec{S}_{x_0,y}\cdot \vec{S}_{x_0+x,y}\rangle|, 
\end{eqnarray}
where $\vec{S}_{x,y}$ is the spin operator on site $i=(x,y)$. $(x_0,y)$ is the reference site indicated by the blue oval shown in Fig.\ref{Fig:DensityProfile} and $x$ is the distance between two sites in $\hat{x}$ direction. Following the same procedure as $A_{cdw}(L_x)$ and $\Phi_{yy}(L_x/2)$, we first extrapolate $F(L_x/2)$ for a given cylinder of length $L_x$ to the limit $m=\infty$, and then perform finite-size scaling as a function of $L_x$. As shown in Fig. \ref{Fig:SC_SS}(b), $F(L_x/2)$ decays exponentially with $L_x$, i.e., $F(L_x/2)\propto e^{-L_x/2\xi_s}$, with the corresponding spin-spin correlation length $\xi_s$ of $8\sim 10$ lattice spacings (Table \ref{Table:KK}). Therefore, we conclude that the spin-spin correlations are short-ranged with a finite gap in the spin sector. Again this is in sharp contrast with the case $t^{\prime}=0$, reflecting the metallic nature of the doped stripes.

Although the spin-spin correlations decay exponentially with $L_x$, it still is dominant over SC correlations up to relatively long distances. To see this, we make a direct comparison between the spin-spin and SC correlations in the same plot in Fig. \ref{Fig:SC_SS} using both double-logarithmic and semi-logarithmic scales. The comparison suggests that relatively large systems, such as $L_x\sim 180$ cylinders, are necessary to see dominant long-range SC correlations. This point stresses the importance of cylindrical length and convergence in prior DMRG studies.

%%============Summary and Discussion==============
\textbf{Discussion:} %
Taken together, our DMRG results indicate the the filling of stripes is a key ingredient that controls the balance between charge and spin-density wave order and superconductivity, with the next-nearest neighbor hopping $t^{\prime}$ being a key tuning parameter to destabilize filled insulating charge stripes. Our results indicate that a route towards stable long-range SC order may lie in mechanisms that perturb the intertwined balance between various predominant correlations. Presumably $t^{\prime}$ alone may not be solely responsible for depopulating filled charge stripes in real materials, as other factors, such as further range hoppings, other orbital contributions, and dynamical lattice effects may also destabilize insulating charge stripes. Answering these open questions may lead to a better understanding of robust SC seen in the cuprates.

%%============Acknowledgement==============
\textbf{Acknowledgement:}
We would like to thank D. J. Scalapino, J. Tranquada, J. Zaanen, B. Moritz, Y. F. Jiang, E. Huang and especially S. Kivelson for insightful discussions. This work was supported by the Department of Energy, Office of Science, Basic Energy Sciences, Materials Sciences and Engineering Division, under Contract DE-AC02-76SF00515. Parts of the computing for this project was performed on the Sherlock cluster.

%%%%%%%%%%%%%%%%%%%%%%%%%%%%%%%%%
%\bibliography{Ref}

%\begin{thebibliography}{5}
%\end{thebibliography}
\bibliographystyle{ieeetr}
%\bibliography{bibs}

%%%%%%%%%%%%%%%%
\appendix 
%\newpage

\begin{center}
\noindent {\large {\bf Supplemental Material}}
\end{center}

\renewcommand{\thefigure}{S\arabic{figure}}
\setcounter{figure}{0}%reset counter
\renewcommand{\theequation}{S\arabic{equation}}%redefine command that creates equation no.
\setcounter{equation}{0}%reset counter

%===============FigS1: Sz and convergence===============
\begin{figure}
  \includegraphics[width=\linewidth]{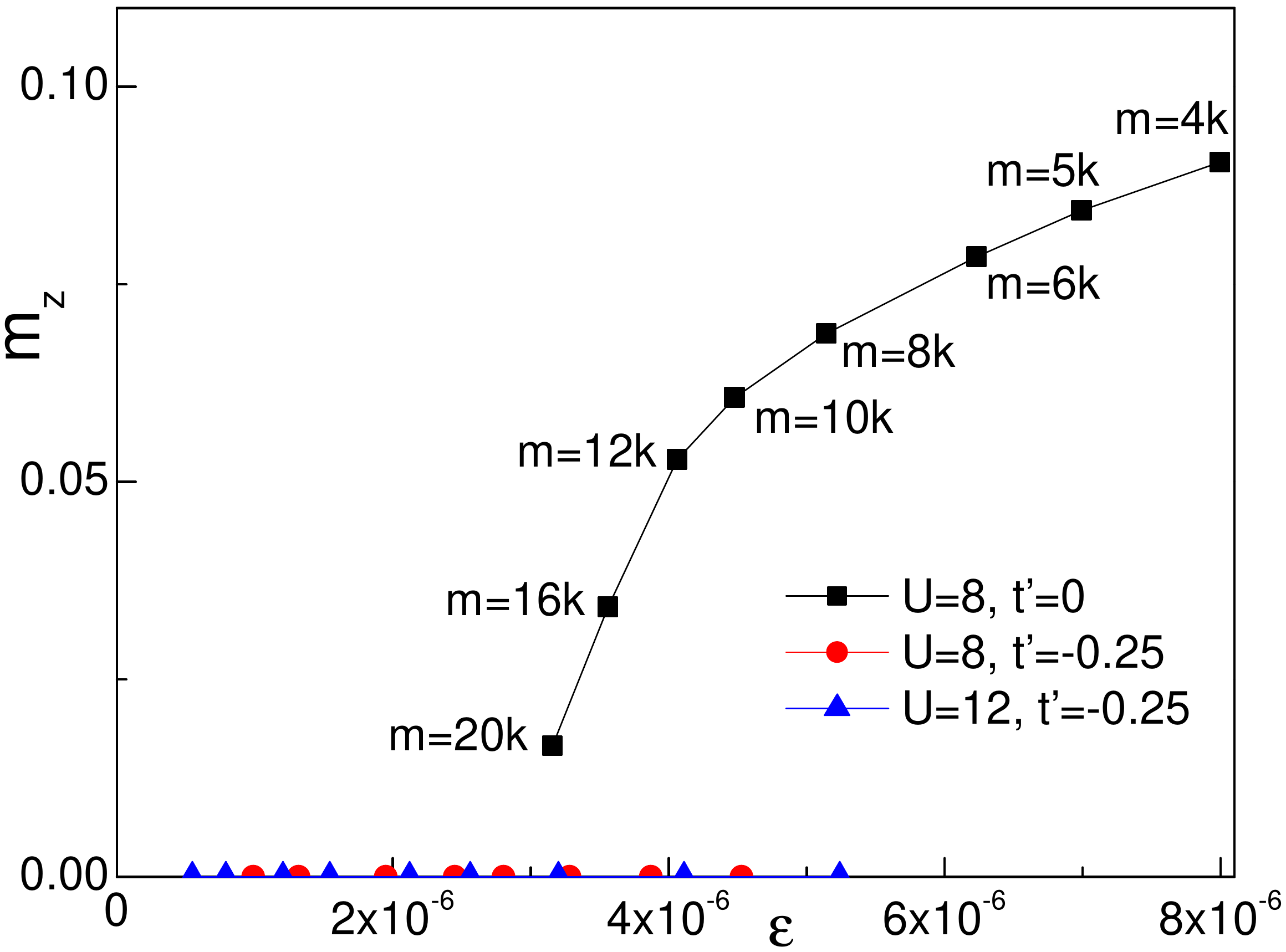}
  \caption{(Color online) Magnetic moment $m_z=\frac{1}{N}\sum_{i=1}^{N}|\langle S^z_i\rangle$ as a function of $m$ or $\epsilon$ of the Hubbard model at doping level $\delta=12.5\%$ on $L_x=64$ cylinders for $t^\prime=0$ and $U=8$ (black squares), $t^\prime=-0.25$ and $U=8$ (red circles) and $U=12$ (blue triangles).}\label{FigS:Sz}
\end{figure}

%%============Numerical convergence==============
\section{I. Numerical convergence}\label{SM:Convergence} %
We have checked the numerical convergence of our DMRG simulations regarding spin rotational symmetry. It is known that in a finite-size system in one dimension or two dimensions, there can be no spontaneous breaking of continuous symmetry. Therefore, the $SU(2)$ spin rotational symmetry of the Hubbard model Hamiltonian cannot be broken in the true ground state. This can be considered as one of the key signatures to determine whether a DMRG simulation has converged to the real ground state.

Our approach to address this issue takes two routes. First, we determine the expectation value of the $z$-component of the spin operator, i.e., $\langle\hat{S}^z_i\rangle$, where $i$ labels the lattice site. Since the ground state is an equal-weight superposition of $|S^z=1/2\rangle$ and $|S^z=-1/2\rangle$ spin states, then $\langle \hat{S}^z_i\rangle=0$ for all sites $i$. A simple measurement of this condition is to define a quantity $m_z=\sum_{i=1}^N |\langle S^z_i\rangle|/N$, which should vanish as the DMRG simulation converges to the true ground state. In all of our DMRG simulations with $t^\prime=-0.25$, we find that $m_z=0$ even when we keep a relatively small number of states, as shown in Fig. \ref{FigS:Sz}, suggesting that our simulations have converged. Unfortunately, for $t^\prime=0$ , in which a much larger number ($m=20000$) of states are kept, a finite $m_z>0$ is obtained, although it decreases rapidly with $m$. Second, the $SU(2)$ spin rotational symmetry requires that the relation $\langle S^x_i S^x_j\rangle$=$\langle S^y_i S^y_j\rangle$=$\langle S^z_i S^z_j\rangle$ holds between two arbitrary sites $i$ and $j$. This relation again is fulfilled in our simulations (not shown), which is contrary to the case of $t^\prime=0$. In addition to spin rotational symmetry, other symmetries including both the lattice translational symmetry in $\hat{y}$ direction and reflection symmetry in $\hat{x}$ direction are also fulfilled. Therefore, we conclude that our simulation for $t^\prime=-0.25$ has converged to the true ground state.

%%=========Further calculation details=========
\section{II. Further calculation details} %
To reliably describe ground state properties, we have explored the role of cylindrical size and boundary effects. In the current study we typically start our calculation with a random state. However, to elucidate the reliability of our results, we also check our calculations by adding a pinning field with the appropriate wavelength to stabilize a CDW state, for example. We find that in all the cases it is sufficient to add the pinning field during the initial sweeps of the calculation and ramp its amplitude to zero in a few subsequent sweeps. This happens only for the smallest number of states that we have considered, i.e., $m=4096$, while for the larger calculations with $m>4096$ it is not necessary to hold a finite (even vanishingly small) pinning field to stabilize the charge stripe pattern. This gives us the same results as we start from a completely random initial state without any pinning field, which undisputedly proves the reliability of our study. Moreover, there is no pinning pair-field to stabilize superconductivity throughout our DMRG calculation.

%==========FigS2: Ground state energy===========
\begin{figure}
  \includegraphics[width=\linewidth]{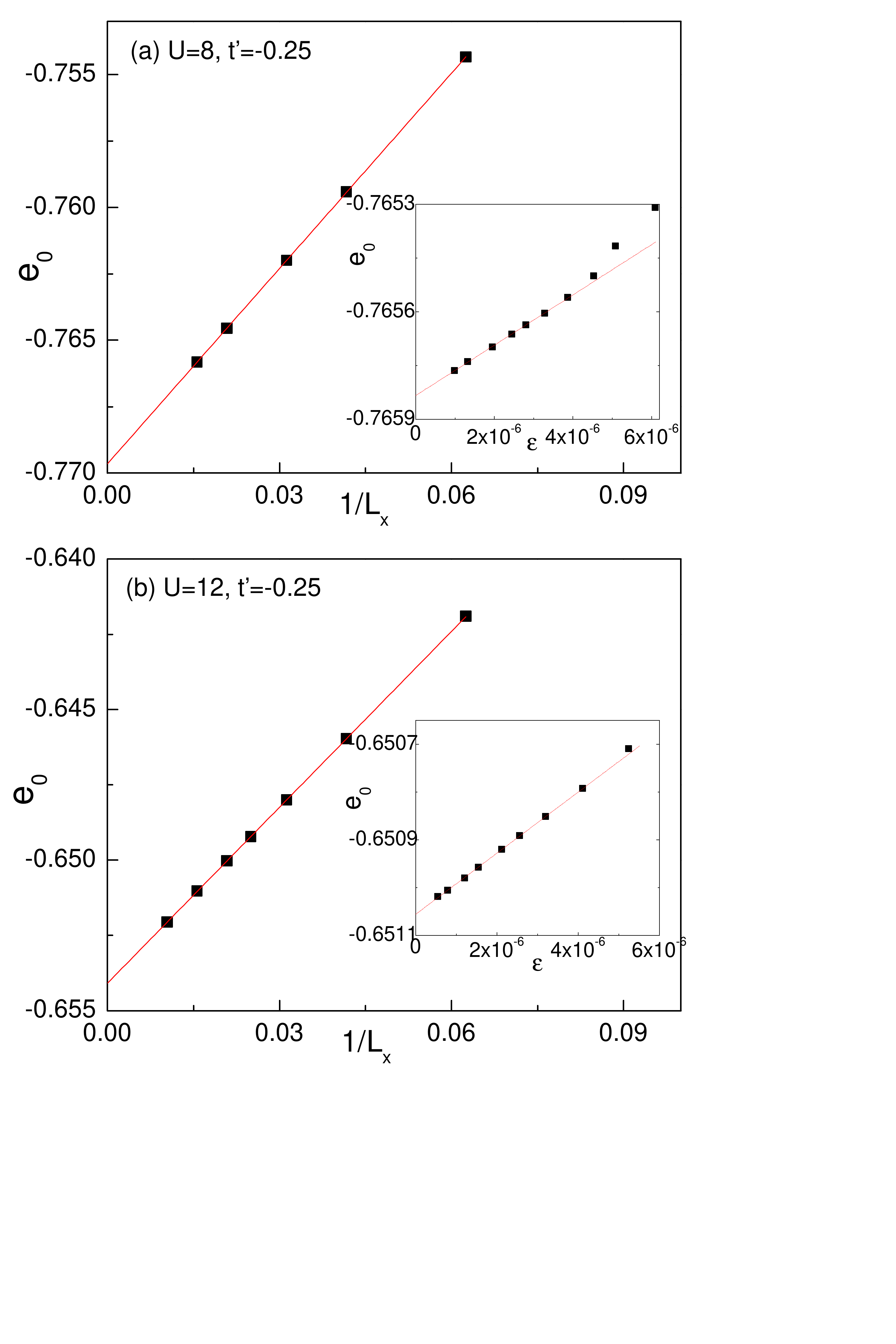}
  \caption{(Color online) Ground state energy per site $e_0$ for $\delta=12.5\%$ as a function of the inverse cylinder length $L_x$ for (a) $U=8$ and (b) $U=12$. Insets: Examples of truncation error $\epsilon$ extrapolation of $e_0$ for $L_x=64$ cylinders at doping level $\delta=12.5\%$ for $U=8$ in (a) and $U=12$ in (b).The red lines show the extrapolation using a linear function.}\label{FigS:EG}
\end{figure}

%=======FigS3: Superconducting correlation function===============
\begin{figure}
  \includegraphics[width=\linewidth]{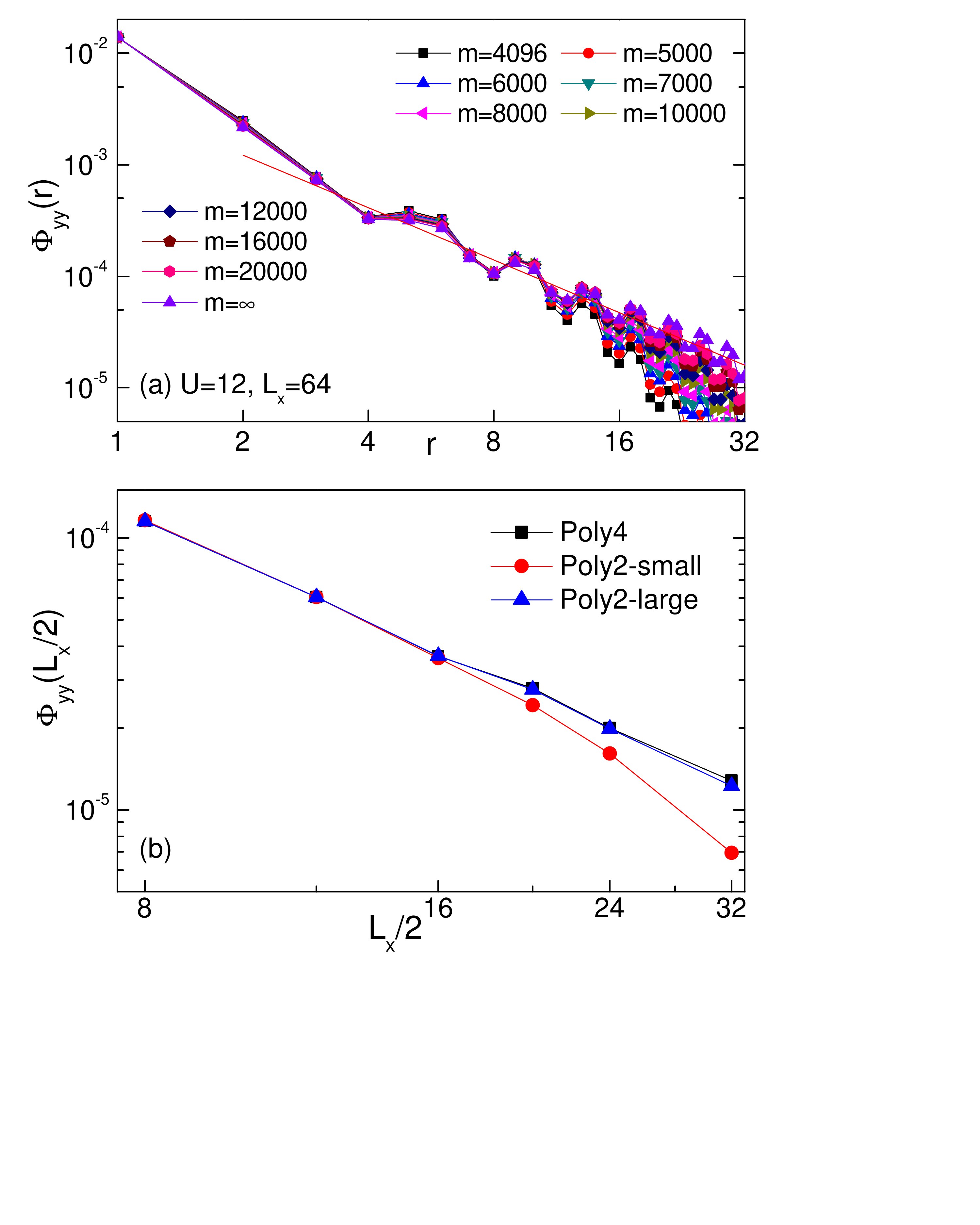}
  \caption{(Color online) (a) SC correlation $\Phi_{yy}(r)$ on a $L_x=64$ cylinder for $\delta=12.5\%$, keeping $m=4096\sim 20000$ states in the double-logarithmic plot, where $r$ is the distance between two Cooper pairs in the $\hat{x}$ direction. The red line represents a power-law fit in the limit $m=\infty$ with $r=1\sim L_x/2$. (b) Extrapolated $\Phi_{yy}(L_x/2)$ using a quartic polynomial (Poly4) and quadratic polynomial (Poly2) function in a double-logarithmic plot (see text).}\label{FigS:SCCor}
\end{figure}

%%=========Ground state energy=========
\section{III. Ground state energy} %
In the insets of Fig. \ref{FigS:EG}, we show examples of truncation error $\epsilon$ extrapolation of the energy per site $e_0=E_0/N$, where $E_0$ is the total energy of a system with $N$ lattice sites, for $L_x=64$ cylinders at doping level $\delta=12.5\%$ and $t^\prime=-0.25$ for $U=8$ (a) and $U=12$ (b). By keeping $m=4096\sim 20000$ number of states, we are able to converge to the true ground state of the system by preserving all symmetries of the Hamiltonian, including the $SU(2)$ spin rotational symmetry, lattice translational symmetry in $\hat{y}$ direction and reflection symmetry in $\hat{x}$ direction. The truncation error extrapolation using a linear function with $m=6000\sim 20000$ gives us $e_0=-0.76583(1)$ for $U=8$ and $e_0=-0.65104(1)$ for $U=12$. The ground state energy $e_0$ of other cylinders can be obtained similarly. Finally, we can obtain accurate estimates of the ground state energies in the long cylinder length limit, i.e., $L_x=\infty$, by carrying out finite-size scaling as a function of the inverse cylinder length. The extrapolation to the limit $L_x=\infty$ for doping level $\delta=12.5\%$ is shown in Fig. \ref{FigS:EG}, in which all energies for cylinder lengths $L_x=16\sim 64$ for $U=8$ and $L_x=16\sim 64$ for $U=12$ fall perfectly onto a linear fit, with a linear regression $R^2$ larger than 99.999\%. This gives an energy $e_0=-0.76965(2)$ for $U=8$, and $e_0=-0.65409(1)$ for $U=12$ in the long cylinder limit $L_x=\infty$. For comparison, we have also obtained the ground state energy $e_0=-0.7661(2)$ for $U=8$ and $t^\prime=0$ in long cylinder limit, which is consistent with previous studies \cite{Zheng_Science_1155}.

%=======FigS4: Friedel oscillation========
\begin{figure}
  \includegraphics[width=\linewidth]{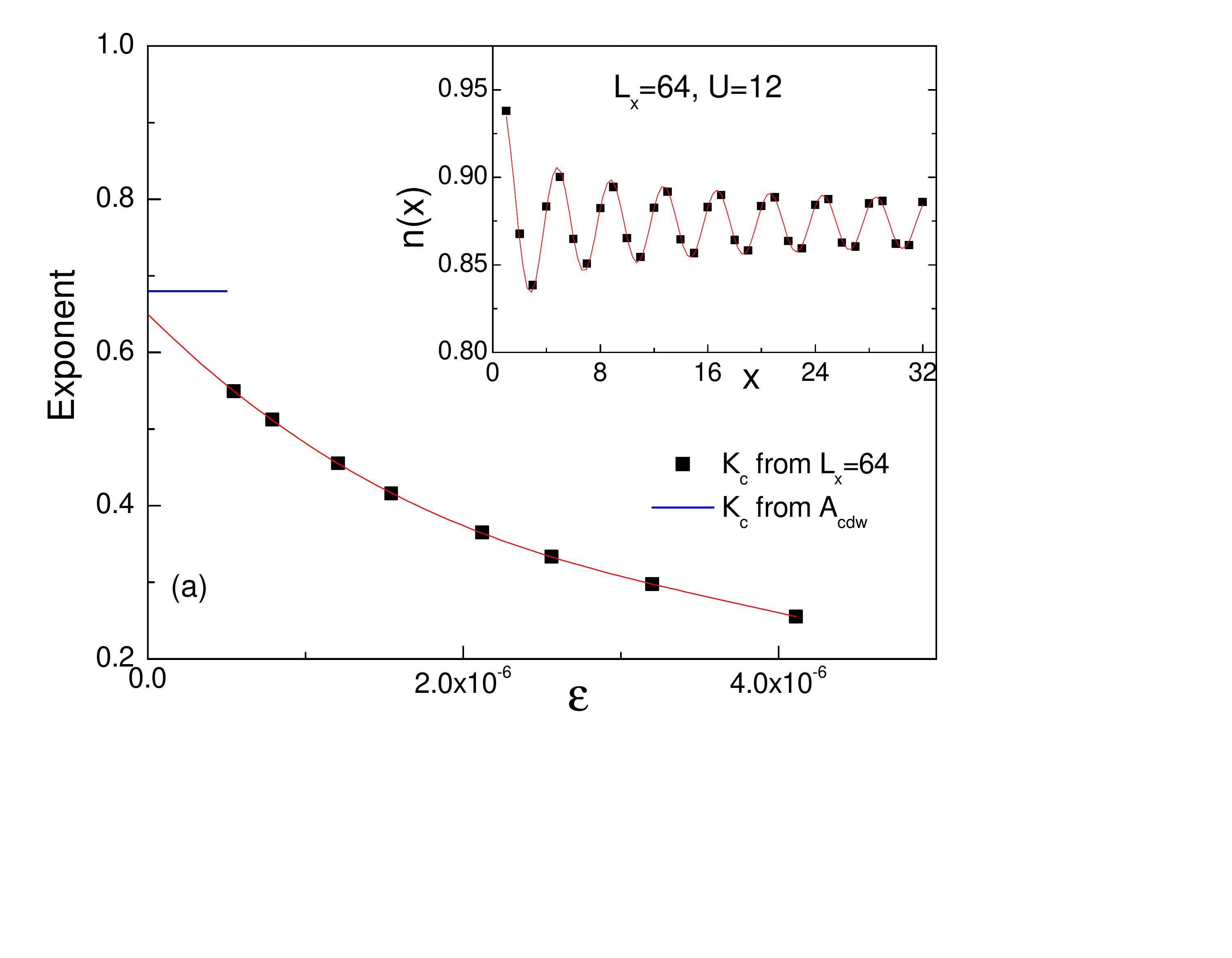}
  \caption{(Color online) Luttinger exponent $K_c$ extracted from the local density profile $n(x)$ with Friedel oscillations on a $L_x=64$ cylinder for $\delta=12.5\%$, by keeping $m=4096\sim 20000$ states. The blue line represents the exponents determined from $A_{cdw}(L_x)$. Inset: Fit of $n(x)$ (solid line) on a $L_x=64$ cylinder using function $n(x)=n_0 + \delta n\ast {\rm cos}(2k_F x + \phi)x^{-K_c/2}$, where $x=1\sim \frac{L_x}{2}$ is the rung index.}\label{FigS:Friedel}
\end{figure}

%%==========Superconducting correlation function===============
\section{IV. Convergence of superconducting correlations }\label{SM:SCCorrelation} %
Fig.\ref{FigS:SCCor} (a) shows the SC pair-field correlation function $\Phi_{yy}(x)$ for $x=1\sim 32$ in $\hat{x}$ direction on a $L_x=64$ cylinder by keeping $m=4096\sim 20000$ states, at doping level $\delta=12.5\%$. The purple triangles label the extrapolated values to the limit $\epsilon=0$, i.e., $m=\infty$, using quartic polynomials, which is consistent with a power-law decay $\Phi_{yy}(x)\propto |x|^{-K_{sc}}$, as indicated by the red solid line.

Fig.\ref{FigS:SCCor} (b) plots the extrapolated $\Phi_{yy}(L_x/2)$ on cylinders of length $L_x=16\sim 64$ at doping level $\delta=12.5\%$, again fitted by different orders of polynomial functions. The black squares label the extrapolated $\Phi_{yy}(L_x/2)$ using a quartic polynomial (Poly4), while the red circles denote the results fitted by a quadratic polynomial, keeping up to $m=10000$ states (Poly2-small). For contrast, the blue triangles represent results fitted by the same quadratic polynomial function but only using 5 data points with the largest number of states (Poly2-large) for each cylinder. From the figure we can clearly see that both Poly4 and Poly2-large fittings are consistent with each other and enough to capture the long distance behavior of the SC pair-field correlation, while the Poly2-small fitting by keeping up to $m=10000$ is not. This may explain the absence of long-range superconductivity in previous DMRG studies.

%=======Luttinger exponent extracted from Friedel oscillation===============
\section{V. Friedel oscillations of the density profile and density-density correlation function}\label{SM:DensityCor} %
Alternatively, the exponent $K_c$ can be extracted by fitting the Friedel oscillation, which is induced by the open boundaries of the cylinder, of the charge density distribution.\cite{White_PRB_2002} In this work, we use $n(x)=n_0 + \delta n\ast {\rm cos}(2k_F x + \phi)x^{-K_c/2}$ to fit the local density profile to extract the Luttinger exponent $K_c$. Here, $\delta n$ is the non-universal amplitude, $\phi$ is a phase shift, $n_0$ is the background density and $k_F$ is the Fermi wavevector. An example is given in the inset of Fig. \ref{FigS:Friedel} for $L_x=64$ cylinder at doping level $\delta=12.5\%$ with rung index $x=1\sim L_x/2$ by keeping $m=20000$ states. The main panel shows the extracted value of $K_c$ from the $L_x=64$ cylinder at the same doping level. In the limit of $m=\infty$, the extracted exponent from $L_x=64$ cylinder is consistent with that determined from $A_{cdw}(L_x)$ (see Fig.2 in the main text).

\end{document}